# Defecton Contribution to the High-Temperature Superconductivity of Hydrides

A.I. Morosov[a]✶ and A.S. Sigov[b]

[a]Moscow Institute of Physics and Technology (National Research University), 9 Institutskiy per., 141700 Dolgoprudny, Moscow Region, Russian Federation

[b]MIREA - Russian Technological University, 78 Vernadsky Ave., 119454 Moscow, Russian Federation

✶ e-mail: mor-alexandr@yandex.ru

## Abstract

We introduce below the concept of defecton, and describe briefly the electron polaron effect, the clustering of defectons, and also the defecton mechanism of superconductivity. It is shown that in the case of high-temperature superconducting metal hydrides, which acquire superconducting properties under pressure of hundreds of gigapascals, this mechanism can make a significant contribution to the Cooper pairing.

## I.   Quantum defects or defectons

For the first time, quantum defects were considered by A.F. Andreev and I.M. Lifshits [1]. They studied the diffusion of impurities in quantum helium crystals and drew attention to the fact that a quantum defect in an ideal crystal is completely delocalized due to tunneling and in fact may be called as a "defecton" quasiparticle. The mean free path of defectons in a helium crystal is determined by their interaction with phonons. A.F. Andreev and I.M. Lifshitz suggested that hydrogen in some metals can also be a quantum impurity. In this case, its mean free path should be determined by collisions with electron excitations. The first consideration of quantum defects in metals and the contribution of defectons to the superconducting pairing was carried out in Ref. [2].

In the case of concentrated metal hydrides close to the stoichiometric composition, quantum defects are vacancies in the sublattice of hydrogen atoms, which is completely filled in the ground state with stoichiometric composition, or hydrogen atoms in the sublattice of interstitials, which is completely free under the same conditions. The first option is implemented at a hydrogen concentration less than the stoichiometric one, and the second option at a concentration higher than the stoichiometric one.

## II. Electron polaron effect

The consideration of the electron-defecton interaction cannot be limited to the Born approximation. Strong renormalizations of the defecton parameters in the metal are a consequence of the electron polaron effect which means the formation of a "coat" of electrons around the defecton. J. Kondo, was the first to pay attention to this [3]. For the sequential calculation of the polaron effect, it is necessary to take into account parquet diagrams in a technique similar to that developed in Refs. [4–6] with the aim of describing the anomalies of the absorption and emission of X-rays in metals. This problem was solved in a series of our works, see the review paper [7]. As an example, we give an expression for the renormalization of the vertex of the electron-defecton interaction $V(\boldsymbol{q})$ ($\boldsymbol{q}$ is the wave vector)

$$V(\boldsymbol{q}) = V_0(\boldsymbol{q}) \left[\frac{E_0}{\max(T,\varepsilon_0)}\right]^g. \tag{1}$$

Here $V_0(\boldsymbol{q})$ is the seed vertex, $E_0$ is the half-width of the electron band, $T$ is the temperature, $\varepsilon_0$ is the defecton band width, and the parameter $g$ is denoted by the integral

$$g = 2\int \frac{d^2k\, d^2k'\, |V_0(\boldsymbol{k}-\boldsymbol{k}')|^2}{(2\pi)^6 |\nabla E(\boldsymbol{k})||\nabla E(\boldsymbol{k}')|}, \tag{2}$$

where the integration is performed over the Fermi surface of electrons and $E(\boldsymbol{k})$ is the electron dispersion law. The value of the parameter $g$ is about 0.1-1.

It can be easily seen that since $E_0 \gg T, \varepsilon_0$, the electron polaron effect enhances the electron-defecton interaction by $\left[\frac{E_0}{\max(T,\varepsilon_0)}\right]^g$ times. The width of the defecton band decreases approximately by the same amount [7].

### III. Defecton clustering

The clustering of hydrogen atoms in dilute metal hydrides at low temperatures prevents the atom band motion. Such a clustering results in either precipitation of a phase with a high hydrogen concentration or the formation of metastable sedentary hydrogen clusters in a metal matrix.

This process inevitability is caused by the alternating character of the long-range interaction between point defects in metal [8]. It is composed of elastic interaction $W_{elas}$ and interaction through Friedel electron-density oscillations $W_{el}$:

$$W(\mathbf{R}) = W_{elas}(\mathbf{R}) + W_{el}(\mathbf{R}), \qquad (3)$$

where $\mathbf{R}$ is the radius vector connecting the defects.

The elastic interaction is determined as follows [9]

$$W_{elas}(\mathbf{R}) = W_1(\mathbf{n})\frac{\Omega}{R^3}, \qquad (4)$$

where $\Omega$ is the unit-cell volume, $\mathbf{n} = \mathbf{R}/R$, and $W_1(\mathbf{n})$ takes positive or negative values, depending on the orientation of vector $\mathbf{n}$ with respect to the crystallographic axes. The $W_1(\mathbf{n})$ value varies from 1 eV for heavy interstitial impurities to $10^{-2}$ eV for two hydrogen atoms in metal [10].

The interaction through Friedel electron-density oscillations can be written for a spherical Fermi surface with radius $k_F$ in the following form [11]

$$W_{el}(\mathbf{R}) = W_2\frac{\Omega}{R^3}\cos(2k_F R), \qquad (5)$$

where $W_2 = $ const $\sim 10^{-2}$ eV. It is as important as the elastic interaction and similarly decreases with increasing distance between defects.

The alternating character of $W(\mathbf{R})$ induces a number of bound states of two defectons or a defecton and a heavy "frozen" impurity. The highest binding energy $W_0$ corresponds to the $R$ of order of interstitial distance. The short-range component of interaction between the defects can change the interaction-energy sign only for several of the smallest $R$ values. Therefore, all mobile point defects in metal (and neutral impurities in insulator) are clusterized with lowering temperature [8, 12]. The characteristic clusterization temperature at low defecton concentrations $x \ll 1$ has the following order of magnitude:

$$T_{cl} \approx W_0/|\ln x|. \qquad (6)$$

Clusters of light and heavy impurities arise in a crystal, in which the concentration of heavy "frozen" impurities is comparable with that of defectons. In some cases, the hydrogen atom occupies one of two equivalent (with respect to the binding energy with the heavy impurity) equilibrium positions at two neighboring interstitial sites. Due to its tunneling between them, a two-level system (TLS) arises. An example of TLS occurrence is the hydrogen capture by heavy impurities of oxygen, nitrogen, or carbon in niobium.

The experimental value of the TLS tunnel matrix element $t \sim 90$ μeV $\sim 1$ K [13] shows that, at normal pressure, $\varepsilon_0$ is usually much less than $T_{cl}$. To realize the band motion of a quantum defect, it is necessary that $2\pi gT \ll \varepsilon_0$ [1]. Since at $T > T_{cl}$ the number of electronic excitations is so large that the motion of the hydrogen atom has not a band character but hopping one, the realization of the defecton mechanism of superconductivity at normal pressure becomes impossible.

### IV. Metal hydrides at high pressure

The discovery of superconductivity in the $H_3S$ compound with the critical temperature $T_c$=203 K at the pressure of 155 GPa [14] attracted attention to the study of metal hydrides. In 2019, superconductivity with $T_c$=250-260 K at the pressure of 170–200 GPa was discovered in the $LaH_{10}$ compound [15, 16]. Formation of high-temperature superconducting phases requires high pressure, which causes a crystal lattice compression and, correspondingly, an exponential increase to values of the order of 100 K of the tunneling matrix element describing the probability of tunneling of the hydrogen atom between equivalent interstitial sites.

The value of the tunnel matrix element can be estimated as

$$t(a) = \text{Ry} \int_{-\infty}^{+\infty} \psi_0\left(x - \frac{a}{2}\right)\psi_0\left(x + \frac{a}{2}\right)dx, \qquad (7)$$

where Ry is the Rydberg energy of 13.6 eV, $\psi_0(x)$ is the wave function of the harmonic-oscillator ground state, and $a$ is the dimensionless distance between the hydrogen atom equilibrium positions at neighboring interstitial sites

$$a = d\sqrt{\frac{m\omega}{\hbar}}, \qquad (8)$$

where $d$ is the distance between the interstitial sites, $m$ is the hydrogen atomic mass, and $\omega$ is the characteristic frequency of optical or local vibrations of this atom. It can be easily seen that

$$t(a) = Ry \exp\left(-\frac{a^2}{4}\right). \qquad (9)$$

Using the value $\hbar\omega$= 160 meV for $\omega$ (which was chosen based on the phonon spectra calculated in [16]), we obtain the $t$ values listed in Table.

**Table.** Dependence of the tunnel matrix element on the interstitial spacing

| $t$, K | $a$ | $d$, Å |
|---|---|---|
| 100 | 5.42 | 0.87 |
| 10 | 6.22 | 1.00 |
| 1 | 6.92 | 1.11 |

The numerical simulation carried out for a number of high-temperature superconducting hydrides shows that $d\lesssim1$ Å in many superconducting phases [17–19] (e.g., 0.841 Å in $SnH_4$ [19]).

Thus, at high pressures, a situation is possible when $\varepsilon_0 > T_{cl}$ and the band motion of defectons takes place.

## V. Defecton contribution to superconductivity

The contribution of defectons to the superconductivity was consistently described in [21]. This contribution is due to the inelastic scattering of electrons from defectons. The corresponding diagrams for the normal and anomalous self-energy parts of the electron Green function are shown in Fig. 1.

On the assumption that the electron‑defecton interaction makes the main contribution to the superconductivity, we have within the weak-coupling approximation ($\lambda_d \ll 1$) the following expression for the critical temperature

$$T_c \approx \varepsilon_0 \exp\left(-\frac{1+\lambda_d}{\lambda_d}\right), \tag{10}$$

where

$$\lambda_d \approx \frac{xgE_0}{\varepsilon_0}. \tag{11}$$

At the applicability limit, the evaluation (10) yields the maximum attainable (based on the defecton superconductivity mechanism) value of $T_c \approx \varepsilon_0$ [22]. This particular fact makes the defecton superconductivity mechanism significant in the high-pressure phases.

## VI. Conclusion

In metal hydrides at high pressure, the width of the defecton zone increases so much that their band motion can be realized at temperature exceeding the clusterization temperature of defects. Under these conditions, the defecton mechanism of superconductivity makes a significant contribution to the Cooper pairing of electrons.

**Fig. 1.** Diagrams of inelastic contribution of defectons to normal (a) and anomalous (b) self-energy parts of the electron Green function. The renormalized defecton Green functions are shown by double dotted lines, normal (a) and anomalous (b) electron Green functions are shown by solid lines, the seed amplitudes of the electron scattering from a defecton are shown by wavy lines, and the total amplitudes of electron scattering from a defecton are shown by wavy lines with a black triangle.

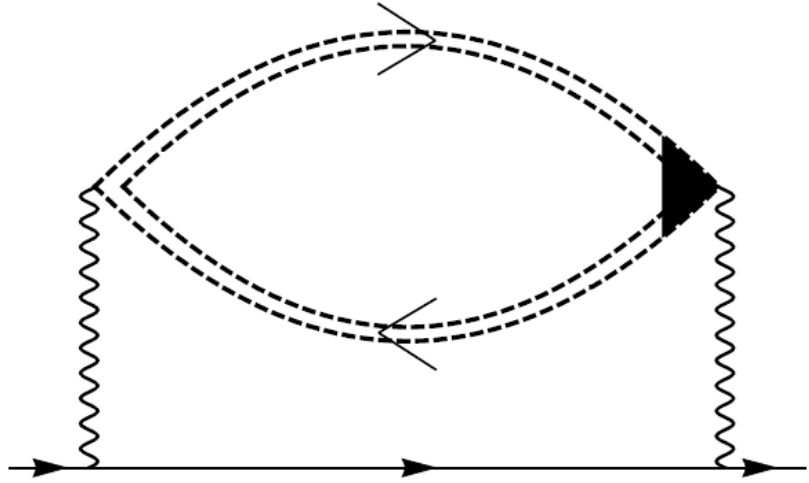

Fig. 1a

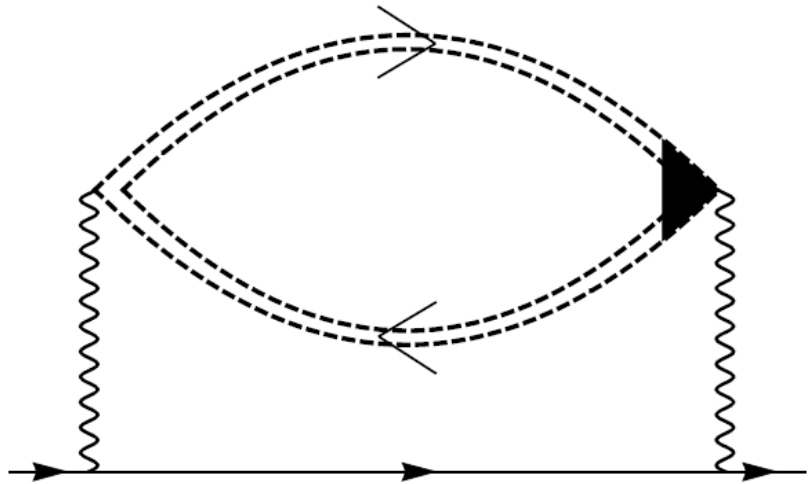

Fig. 1b